\documentclass[onecolumn,showpacs,superscriptaddress, preprintnumbers,amsmath,amssymb,nofootinbib]{revtex4}
\usepackage{dcolumn}
\usepackage{bm}
\usepackage{tabularx}
\usepackage{array}
\usepackage{graphics}
\usepackage{graphicx}
\usepackage{psfrag}
\usepackage{epsfig}
\usepackage{amsmath}
\usepackage{amssymb}
\usepackage{verbatim}
\usepackage{color}
\usepackage{psfrag}
\usepackage{textcomp}
\usepackage[usenames,dvipsnames]{xcolor}

\usepackage{verbatim}

\newcommand{\bea}{\begin{eqnarray}}
\newcommand{\eea}{\end{eqnarray}}
\newcommand{\beq}{\begin{equation}}
\newcommand{\eeq}{\end{equation}}

\newcommand{\pdir}{p\kern -5.2pt\raise 0.2ex\hbox {/}}
\newcommand{\vdir}{v\kern -5.75pt\raise 0.15ex\hbox {/}}
\newcommand{\kdir}{k\kern -5.75pt\raise 0.15ex\hbox {/}}
\newcommand{\epsdir}{\epsilon\kern -5.0pt\raise 0.15ex\hbox {/}}
\newcommand{\bvdir}{\bar{v}\kern -5.75pt\raise 0.15ex\hbox {/}}
\newcommand{\Ddir}{D\kern -7.75pt\raise 0.20ex\hbox {/}}
\newcommand{\Adir}{A\kern -7.75pt\raise 0.20ex\hbox {/}}
\newcommand{\ldir}{l\kern -5.0pt\raise 0.2ex\hbox{/}}
\newcommand{\varepsdir}{\varepsilon\kern -5.5pt\raise 0.15ex\hbox{/}}

\newcommand{\lgl}{\langle}
\newcommand{\rgl}{\rangle}

\def\elematrice#1#2#3{\lgl#1|#2|#3\rgl}

\makeatother

\begin{document}
\preprint{\tt LPT Orsay 13-22}
\preprint{\tt TCD-MATH 13-04}
\preprint{\tt DESY 13-069}
\vspace*{22mm}
\title{On the $B^{*'}\to B$ transition}

%

\author{Beno\^it~Blossier}
\affiliation{Laboratoire de Physique Th\'eorique (B\^at. 210), Universit\'e Paris Sud,  Centre d'Orsay, 
91405 Orsay-Cedex, France}
\author{John~Bulava\footnote[1]{On a leave of absence from CERN}}
\affiliation{School of Mathematics, Trinity College, Dublin 2, Ireland}
\author{Michael~Donnellan}
\affiliation{DESY, Platanenallee 6, D-15738 Zeuthen, Germany}
\author{Antoine~G\'erardin}
\affiliation{Laboratoire de Physique Th\'eorique (B\^at. 210), Universit\'e Paris Sud,  Centre d'Orsay,
91405 Orsay-Cedex, France}


\begin{abstract}
We present a first $N_{\rm f}=2$ lattice estimate of the hadronic coupling 
$g_{12}$ which parametrises the strong decay of a radially excited $B^*$ meson 
into the ground state $B$ meson at zero recoil. We work in the static limit of  Heavy Quark Effective Theory (HQET) and solve  
a Generalised Eigenvalue Problem (GEVP), which is necessary for the extraction 
of excited state properties. After an extrapolation to the continuum limit 
and a check of the pion mass dependence, we obtain $g_{12} = -0.17(4)$. 

\end{abstract}

\pacs{12.38.Gc, 13.20.He.}
\maketitle

\section{\label{Introduction}Introduction}

\begin{figure*}[t!]
\epsfxsize9.2cm\epsffile{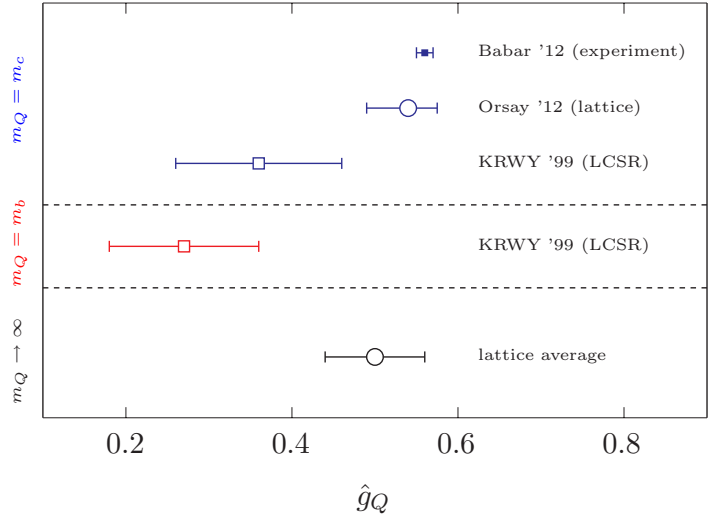}\\
\caption{\label{figg} Experimental measurement \cite{Godang:2013im}, lattice 
computations \cite{OhkiPY,BecirevicYB,BulavaEJ,DetmoldGE,BecirevicPF} and 
sum rules estimates 
\cite{KhodjamirianHB}
of $\hat{g}_c$, $\hat{g}_b$ and $\hat{g} \equiv \hat{g}_\infty$.
We have performed a weighted average of recent $\hat{g}$ lattice results
at ${\rm N_f}=2$ with respect to the error quoted in \cite{OhkiPY,BecirevicYB,
BulavaEJ,DetmoldGE}.}
\end{figure*}

Questions have been raised recently on the poor handling of excited states in 
the analyses of experimental data and their comparison with theoretical
predictions, particularly in the case of heavy-light $B$ and $D$ mesons \footnote{The quantum numbers 
of the low-lying meson (H) are listed in Table \ref{BDspectrum}.}.
For instance, it has been advocated that the  
$\sim$ 3$\sigma$ discrepancy observed between exclusive and inclusive estimates of the CKM matrix element $V_{cb}$ 
might be reduced if the transition $B \to D'$ were large. This attractive 
hypothesis implies a suppression of the
$B \to D^{(*)}$ hadronic form factors, as a study in the OPE
formalism suggests~\cite{GambinoRD}. On the other hand, it 
has been argued that the light-cone sum rule determination of the 
$g_{D^{*}D\pi}$ coupling, which parametrises the $D^{*} \to D \pi$ decay, 
likely fails to reproduce the 
experimental measurement unless one explicitly includes
the contribution from the first radial excited $D^{(*)'}$ state on the 
hadronic side of the three-point Borel sum rule~\cite{BecirevicVP}. Comparison 
with sum rules is of particular importance because the heavy mass dependence 
of $\hat{g}_Q\equiv \frac{g_{H^*H\pi}f_\pi}{2\sqrt{m_H m_{H^*}}}$ deduced from 
recent lattice simulations~\cite{OhkiPY,BecirevicYB,BulavaEJ,DetmoldGE,
BecirevicPF,BernardoniIP} and 
experiment~\cite{Godang:2013im} seems much weaker than expected from analytical 
methods~\cite{KhodjamirianHB}, as shown in Figure~\ref{figg}.
\begin{table}[b|]
\begin{center}
\begin{tabular}{|c|c|c|c|c|}
\hline
$J^P$	&	ground state		&	radial excitation	\\ 
\hline 
$0^{-}$	&	$H$	&	$H^{'}$	\\
$1^{-}$	&	$H^{*}$	& $H^{*'}$	\\ 
\hline 
$0^{+}$	&	$H_0^{*}$	&$H_0^{*'}$	\\
$1^{+}$	&	$H_1^{*}$	&$H_1^{*'}$	\\ 
\hline 
\end{tabular}
\end{center}
\caption{\label{BDspectrum} Spectroscopy of the lowest states of heavy-light $H \equiv$ $B$ and $D$ mesons.}
\end{table} 

Techniques have been developed to study excited states of mesons using lattice 
QCD~\cite{BulavaNP}, especially to extract the spectrum~\cite{BurchCC, BlossierVZ,MohlerKE, MahbubRM}. Similar 
techniques can now be applied to  
three-point correlation functions to perhaps illuminate the 
phenomenological issues discussed above. 
In this letter we will report on the lattice computation of 
$g_{12}\equiv \langle B^{*'}|{\cal A}_{i}|B\rangle$ in 
the static limit of HQET, where
${\cal A}_{i}$ is the axial vector bilinear of light quarks and 
$B^{*'}$ is
polarised along the $i$th direction. As a by-product of our work, we will also 
report on the computation of $g_{11} \equiv \langle B^{*}|{\cal A}_{i}|B\rangle$ and 
$g_{22}\equiv \langle B^{*'}|{\cal A}_{i}|B'\rangle$.

The Heavy Quark Symmetry of leading order HQET is well suited 
for our qualitative study. As the spectra of excited $B$ and $B^*$ mesons are 
degenerate, it is enough to
solve a single Generalized Eigenvalue Problem (GEVP) while degrees of 
freedom $\sim$ $m_b$, that are somehow irrelevant for the
dynamics of the cloud of light quarks and gluons that governs the process we 
examine, are integrated out. The plan of the letter is the following: in 
Sec.~\ref{sec2} we 
describe our approach while in Sec.~\ref{sec3} we present our lattice set-up 
and discuss results before concluding in Sec.~\ref{sec4}.

\section{\label{sec2}Extraction of $\langle B^{*'}|{\cal A}_{i}|B\rangle$}

The transition amplitude of interest is parametrised by
\bea\nonumber
\langle B^{*'}(p',\epsilon_\lambda)|{\cal A}^{\mu}|B(p)\rangle
&=&2m_{B^{*'}}A_0(q^2)
\frac{\epsilon^{(\lambda)}\cdot q}{q^2} q^\mu
+(m_B + m_{B^{*'}}) A_1(q^2)\left(\epsilon^{(\lambda)\,\mu} - 
\frac{\epsilon^{(\lambda)}\cdot q}{q^2}q^\mu \right)\\
&+&A_2(q^2) 
\frac{\epsilon^{(\lambda)}\cdot q}{m_B + m_{B^{*'}}} \left[(p_B+p_{B^{*'}})^\mu + 
\frac{m^2_B - m^2_{B^{*'}}}{q^2}q^\mu\right]\,,
\eea
with $q=p'-p$. In the zero recoil kinematic configuration where $\vec{p}=\vec{p}'=\vec{0}$, 
one has $q^2_{\rm max}=(m_{B^{*'}}-m_B)^2$ so that 
\beq
\langle B^{*'}(p',\epsilon_\lambda)|{\cal A}^{i}|B(p)\rangle
=(m_B + m_{B^{*'}}) A_1(q^2_{\rm max}) \epsilon^{(\lambda)\,i}\,.
\eeq
At that stage it is useful to introduce the HQET normalisation of states:
$|H \rangle = \sqrt{2 m_H} | H \rangle_{\rm HQET}$, with 
$\langle H(p) | H(p') \rangle=
2E(p) \delta^3(\vec{p} - \vec{p}')$:
\beq
\langle B^{*'}(p',\epsilon_\lambda)|{\cal A}^{i}|B(p)\rangle_{\rm HQET}
=\frac{m_B + m_{B^{*'}}}{2\sqrt{m_B m_{B^{*'}}}} A_1(q_{\rm max}^2) \epsilon^{(\lambda)\,i}\,.
\eeq
In the static limit we are left with 
$\langle B^{*'}(p',\epsilon_\lambda)|{\cal A}^{i}|B(p)\rangle_{\rm HQET}
=A_1(q_{\rm max}^2) \epsilon^{(\lambda)\,i}$. Choosing the quantization axis 
along the $z$ direction and the polarisation vector 
$\epsilon^\mu(0)=\left(\begin{array}{c}0\\0\\0\\1\\\end{array}\right)$, with 
the metric (+,-,-,-),
we get finally $A_1(q^2_{\rm max})=\langle B^{*'}(p',\epsilon_0)|{\cal A}^{3}|B(p)\rangle_{\rm HQET}$.
Of course, extracting $g_{11} \equiv \hat{g}$ and $g_{22}$ is similar, except 
that the relevant axial form factors are defined at $q^2=0$.

GEVP methods \cite{MichaelNE, LuscherCK, BlossierKD} are a very efficient tool 
to study excited states on the lattice.
We consider $N \times N$ matrices of two-point correlation functions together with the corresponding matrices of three-point correlation functions 
$C^{(')(2)}_{ij}(t)\equiv 
\lgl O^{(')}_i(t)O^{(')\dag}_j(0)\rgl$ and $C^{(3)}_{ij}(t,t_s)\equiv 
\lgl O'_i(t_s)O_\Gamma(t)O^\dag_j(0)\rgl$, where $i,j$ represent different wave 
functions and Dirac structures with quantum numbers generically denoted $(h)$. More explicitly, the $O_{i}$ are interpolating fields 
of pseudoscalar 
static-light mesons, the $O'_{i}$ interpolating
fields of vector static-light mesons and $O_\Gamma$ the axial vector 
light-light bilinear of quarks.

In HQET the spectral decomposition reads
$C^{(')(2)}_{ij}(t) = \sum_n \psi^{*(h^{(')})}_{ni} \psi^{(h^{(')})}_{nj}
e^{-E_n t}$, $\psi^{(h^{(')})}_{ni} = 
\elematrice{M^{(h^{(')})}_n}{\hat{O}^{\dag(')}_i}{0}$.
The purpose of solving GEVP is to construct quantities which tend toward 
the desired excited state properties asymptotically in time. 
In practice we solve 
\beq
\sum_j C^{(2)}_{ij}(t) v^{(n)}_j(t,t_0)=\sum_j \lambda^{(n)}(t,t_0)
C^{(2)}_{ij}(t_0) v^{(n)}_j(t,t_0),\quad
\lambda^{(n)}(t,t_0)=e^{-E^{\rm eff}_n(t,t_0) (t-t_0)}\,.
\eeq
We will use two ratio methods, GEVP and sGEVP, to extract the matrix element 
$M_{mn}\equiv \elematrice{M^{(h)}_n}{\hat{O}_\Gamma}{M^{(h')}_m}$. Those ratios 
converge quickly as the contribution of higher 
excited states is strongly suppressed~\cite{BulavaYZ}\footnote{We give in the Appendix a hint of the proof
of the $t$ behaviour of $R^{\rm sGEVP}_{mn}(t)$, as it was not discussed in 
detail in~\cite{BulavaYZ}.} and read:
\bea\nonumber
R^{\rm GEVP}_{mn}(t,t_{s})&=&
\frac{\langle v^{(n)}(t_s-t,t_0),C^{(3)}_{\Gamma}(t,t_s)v^{(m)}(t,t_0)\rangle
\lambda^{(m)}(t_0+a,t_0)^{-t/2}
\lambda^{(n)}(t_0+a,t_0)^{t-t_s/2}}
{\sqrt{\langle
v^{(n)}(t_s-t,t_0),C^{(2)}(t_s-t)
v^{(n)}(t_s-t,t_0)\rangle\langle
v^{(m)}(t,t_0),C'^{(2)}(t)
v^{(m)}(t,t_0)\rangle}}\\
\label{ratiogevp}&\stackrel{t/a \gg 1, (t_s-t)/a \gg 1}{\sim}&
M_{mn}+{\cal O}(e^{-\Delta_{N+1,m}t}, e^{-\Delta_{N+1,n}(t_{s}-t)})\,,
\eea
\beq\nonumber
\Delta_{N+1,n}=E_{N+1}-E_{n}\,, \quad \langle a, b\rangle \equiv \sum_i a_i b_i\,.
\eeq

\bea\nonumber
R^{\rm sGEVP}_{mn}(t)&=&\partial_{t} \left[\frac{\lgl v^{(m)}(t,t_{0}), 
[K^{mn}(t,t_{0})/\lambda^{(n)}(t,t_{0})
-K^{mn}(t_{0},t_{0})] v^{(n)}(t,t_{0})\rgl}{\sqrt{\lgl v^{(m)}(t,t_{0}), 
D^{mn}(t,t_{0}) v^{(m)}(t,t_{0})\rgl}
\sqrt{\lgl v^{(n)}(t,t_{0}), C^{(2)}(t_{0}) v^{(n)}(t,t_{0})\rgl}}\right]\\
\label{ratiosgevp}&\stackrel{t/a \gg 1, (t_{s}-t)/a \gg 1}{\sim}&
M_{mn}+{\cal O}(\Delta t e^{-\Delta t_0})\,,
\eea
\begin{align*}\nonumber
K^{mn}_{ij}(t,t_{0})=\sum_{t_{1}} e^{-(t-t_1)\Sigma^{mn}(t,t_{0})} C^{(3)}_{ij}(t_{1},t),
\quad D^{mn}_{ij}(t,t_{0})=e^{-t\Sigma^{mn}(t,t_{0})} C'^{(2)}_{ij}(t),\\
\end{align*}

\vspace{-1cm}
\beq\nonumber
\Sigma^{mn}(t,t_{0})=E^{\rm eff}_{n}(t,t_{0}) - E^{\rm eff}_{m}(t,t_{0}),\quad
a E^{\rm eff}_{n}(t,t_{0}) = \ln \left(\frac{\lambda^{(n)}(t,t_{0})}{\lambda^{(n)}(t+a,t_{0})}\right).
\eeq

In the appendix, we have calculated the time dependence of the corrections in 
$R^{\rm sGEVP}_{mn}(t)$ to first order in $\epsilon$, where 
\beq\nonumber
C^{(2)}_{ij}(t) = C_{ij}^{(2,0)}(t) + \epsilon C_{ij}^{(2,1)}(t) =  
\sum_{n=1}^N e^{-E_nt} \psi_{ni} \psi_{nj}  
+  \sum_{n=N+1}^{\infty} e^{-E_nt} \psi_{ni} \psi_{nj},
\eeq
\beq\nonumber
R_{mn}^{\rm sGEVP}=M_{mn} + \epsilon R_{mn}^{\rm sGEVP, 1}
\eeq
We have found that for $n > m$ the dominant contribution to 
$\epsilon R^{{\rm sGEVP, 1}}_{mn}$ is  $t e^{-(E_{N+1}-E_n) t}$ 
and for $n < m$ the leading contribution is in $e^{-(E_{N+1}-E_m)t}$.

The global phase is 
fixed by imposing the positivity of the `decay constant' $f_{M^{(h)}_n} \equiv 
\elematrice{M^{(h)}_n}{O^\dag_L}{0}=
\frac{\sum_i C^{(2)}_{Li}(t)v^{(n)}_i(t,t_0) \lambda^{(n)}(t_0+a,t_0)^{-t/2}}
{\sqrt{\langle v^{(n)}(t,t_0),C^{(2)}(t)v^{(n)}_j(t,t_0)\rangle}}$, where $L$ 
refers to some local interpolating field.
\begin{figure*}[t!]
\epsfig{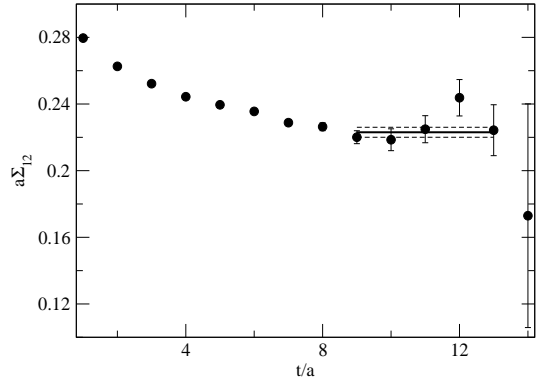}\\
\caption{\label{figsigma12} Plateau of $E_2 - E_1$ for the CLS ensemble E5.}
\end{figure*}

\section{\label{sec3}Lattice results}

We have performed measurements on a subset of the ${\rm N_f}=2$ CLS lattice 
ensembles, which employ the plaquette gauge action and 
non-perturbatively ${\cal O}(a)$ improved 
Wilson-Clover fermions. The parameters of the ensembles used in this work 
are collected in Table~\ref{tabsim}. Three lattice spacings 
($0.05\,\mathrm{fm} \lesssim a \lesssim 0.08\,\mathrm{fm}$) are
considered with pion masses in the range $[310\,\mathrm{MeV},\, 440\,\mathrm{MeV}]$. The static-light correlation functions employ the 
`HYP2' discretization of the static quark 
action~\cite{HasenfratzHP,DellaMorteYC} and stochastically estimated 
all-to-all light quark propagators 
with full time dilution~\cite{FoleyAC}. A single fully time-diluted stochastic 
source has been used on each gauge configuration, except for the ensemble E5 
where we have four stochastic 
sources for each gauge configuration. We use interpolating fields for 
static-light mesons of the so-called Gaussian smeared-form~\cite{GuskenAD}
\beq\nonumber
O_i = \bar{\psi}_h \Gamma\,(1 + \kappa_G a^2 \Delta)^{R_i} \psi_l\,,
\eeq
where $\kappa_G=0.1$ is a hopping parameter, $R_{i}$ is the number of 
applications of the operator $(1 + \kappa_G a^2 \Delta)$, and $\Delta$ the 
gauge-covariant 3-D Laplacian constructed from three-times APE-blocked 
links~\cite{AlbaneseDS}. $R_{i}$ is chosen such that the radius 
$r_{i} \equiv 2a \sqrt{\kappa_G R_i}$ 
of the ``wave-function'' is smaller than $0.6\,\mathrm{fm}$. On each 
ensemble we have estimated the statistical error from a jackknife procedure.

\begin{figure*}[t!]
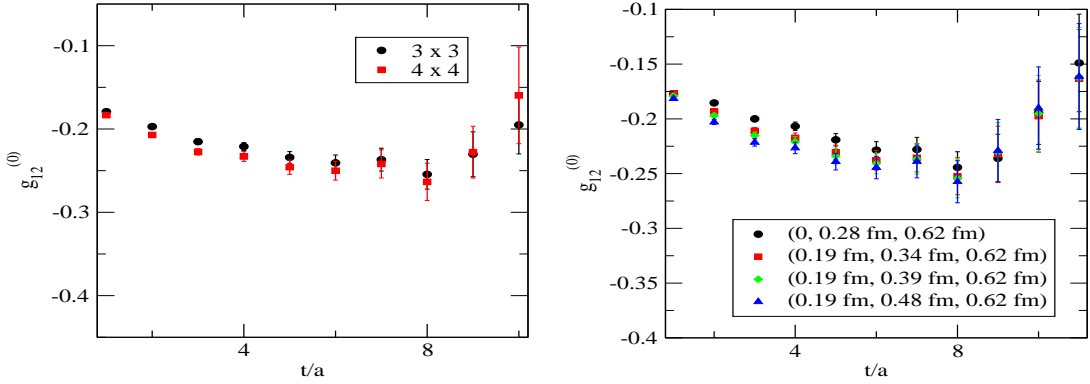

\includegraphics*[width=7cm, height=5cm]{g3344.eps} \quad
\includegraphics*[width=7cm, height=5cm]{ginterpfields.eps}\\
\caption{\label{fignfields} Dependence of bare $g_{12}$ on the size of the GEVP (left) and on the radius of wave functions (right) for the CLS ensemble E5.}
\end{figure*}

In order to reduce the statistical
uncertainty in ratio (\ref{ratiosgevp}), we have taken 
the asymptotic value of the energy splittings 
$\Sigma^{mn}_{\infty}=E_n - E_m$. We have shown in Figure~\ref{figsigma12} an 
example plateau for 
$\Sigma^{12}_{\infty}$. In addition we have set $t_s$ to $2t$ in (\ref{ratiogevp}). We have 
solved both $3 \times 3$ and $4 \times 4$ GEVP systems and checked the stability of the results when the local 
operator is included, as shown in Figure \ref{fignfields}. Hereafter we will
present results for a $3 \times 3$ matrix of correlators with values of 
$r_i \equiv
\{0.19\,\mathrm{fm},\,0.39\,\mathrm{fm},0.62\,\mathrm{fm}\}$.
To check the dependence on $t_0$, to which
the contribution from higher excited states is sensitive, we have both 
fixed it at a small value (typically, $2a$) and let
it vary as $t-a$.

\begin{table}[b|]
\begin{center}
\begin{tabular}{|c|c|c|c|c|c|c|}
\hline
CLS label&$\beta$&$L^3\times T$&$\kappa$&a [fm]&$m_\pi$ [MeV]&\# of cnfgs\\
\hline
A5&5.2&$32^3\times 64$&0.13594&0.075&330&500\\
\hline
E5&5.3&$32^3\times 64$&0.13625&0.065&435&500\\
F6&&$48^3 \times 96$&0.13635&&310&600\\
\hline
N6&5.5&$48^3 \times 96$&0.13667&0.048&340&400\\
\hline
\end{tabular}
\end{center}
\caption{\label{tabsim} Parameters of the simulations.}
\end{table} 

\begin{figure*}[t!]
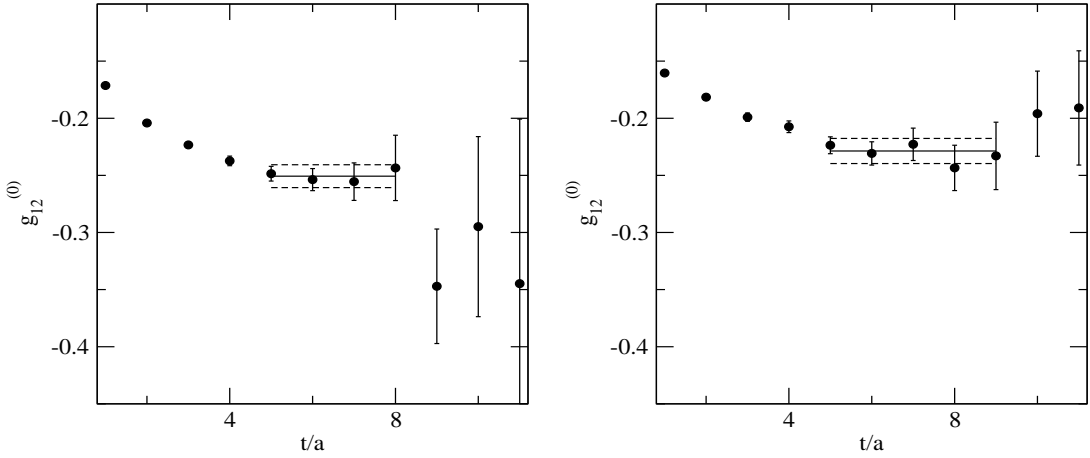

\includegraphics*[width=7cm, height=6cm]{g12GEVP.eps} \quad
\includegraphics*[width=7cm, height=6cm]{g12sGEVP.eps}\\
\caption{\label{figgevpsgevp} Plateaus of bare $g_{12}$ extracted by GEVP 
(left) and sGEVP (right) for the CLS ensemble E5.}
\end{figure*}

 Though the uncertainty is a bit larger, we have confirmed the finding 
 by \cite{BulavaYZ} that
using sGEVP (\ref{ratiosgevp}) seems beneficial compared to the standard GEVP 
approach (\ref{ratiogevp}) 
to more strongly suppress contamination from higher excited states in the 
hadronic matrix element
we measure. As illustrated in Figure~\ref{figgevpsgevp}, plateaux obtained from the GEVP and sGEVP are compatible: -0.25(1) for
GEVP and -0.23(2) for sGEVP, with one additional point in the plateau of the 
sGEVP. Therefore, in the following we give results using 
the sGEVP only.
 
\begin{figure*}[b!]
\includegraphics*[width=7cm, height=6cm]{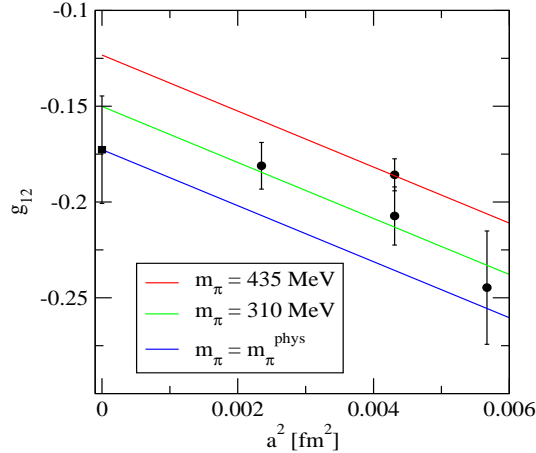}\\
\caption{\label{figgextr} Continuum and chiral extrapolation of $g_{12}$.}
\end{figure*}

 After applying a non-perturbative procedure to renormalise the axial 
 light-light current~\cite{DellaMorteXB,FritzschWQ}, we are ready to 
 extrapolate to the continuum limit.
Inspired by Heavy Meson Chiral Perturbation Theory at leading 
order~\cite{CasalbuoniPG,BurdmanGH} and
due to the ${\cal O}(a)$ improvement of the three-point correlation functions 
 (the improved part of the axial current, $ac_{A} \partial_{i} {\cal P}$, is absent 
at zero momentum),
we apply two fit forms:
\bea\label{fit1}
g_{12}&=&C_0 + (a/a_{\beta=5.3})^2 C_1 + (m_\pi / m^0_\pi)^2 C_2\,,\\
\label{fit2}
g_{12}&=&C'_0 + (a/a_{\beta=5.3})^2 C'_1\,.
\eea
We show in Figure~\ref{figgextr} the continuum extrapolation (\ref{fit1}) of $g_{12}$. 
We observe quite large cut-off effects ($\sim 30\%$ at
$\beta=5.3$), it is thus crucial to have several lattice spacings. We obtain 
finally, using (\ref{fit1}) as the best estimate of the central value,

\beq\label{result}
g_{12}=-0.17(3)(2)\,,
\eeq
where the first error is statistical, 
and the 
second error corresponds to
the chiral uncertainty that we evaluate from the discrepancy between
(\ref{fit1}) and (\ref{fit2}).
 We collect in Table~\ref{tabg12} the value of $g_{12}$ at each lattice point 
 and at the physical point as well as the fit parameters for 
 (\ref{fit1}) and (\ref{fit2}).
\begin{table}
\begin{center}
\begin{tabular}{|c|c|}
\cline{2-2}
\multicolumn{1}{l|}{}&$g_{12}$\\
\hline
A5&-0.245(29)\\
E5&-0.186(8)\\
F6&-0.207(15)\\
N6&-0.181(12)\\
\hline
physical point&-0.173(28)(18)\\
\hline
\end{tabular}
\quad
\begin{tabular}{|c|c|c|}
\cline{2-3}
\multicolumn{1}{l|}{}&fit (\ref{fit1})&fit (\ref{fit2})\\
\hline
$C_{0}$&-0.178(29)&-0.155(26)\\
$C_{1}$&-0.063(32)&-0.040(29)\\
$C_{2}$&0.0053(29)&-\\
\hline
\end{tabular}
\end{center}
\caption{\label{tabg12} Value of $g_{12}$ at the lattice points and at the 
physical point (left)
as well as the fit parameters of Eq. (\ref{fit1}) and (\ref{fit2}) (right).}
\end{table}
\begin{table}
\begin{center}
\begin{tabular}{|c|c|}
\cline{2-2}
\multicolumn{1}{l|}{}&$a\Sigma_{12}$\\
\hline
A5&0.255(8)\\
E5&0.222(8)\\
F6&0.216(12)\\
N6&0.173(7)\\
\hline
\end{tabular}
\end{center}
\caption{\label{tabsigma} Mass splitting $\Sigma_{12}$ in lattice
units. The error we quote is the discrepancy between plateaux that we extract for 
different time ranges \{$[t_{\rm min},\,t_{\rm max}]$, $[t_{\rm min}\pm 0.2 r_0,\,
t_{\rm max}\pm 0.2 r_0]$\}, where the Sommer scale $r_{0}$ \cite{SommerCE} is about 
$0.5\, \mathrm {fm}$ \cite{FritzschWQ}.}
\end{table}

In simulations with light dynamical quarks, the onset of multi-hadron
thresholds due to the emission of pions must be considered when examining
excited $B$ meson properties. Such thresholds significantly complicate the
extraction of hadron-to-hadron matrix elements from the two- and three-point
correlation functions considered here. However with the $L < 3\, \mathrm{fm}$
volumes in this work, the $P$-wave decay $B^{*'}(\vec{0}) \to B(\vec{p}) 
\pi(-\vec{p})$ is
kinematically forbidden. The $S$-wave decay $B^{*'} \to B^*_1 \pi$ is potentially more
dangerous. Examining the mass splittings $\Sigma_{12}$ in Table \ref{tabsigma}, we
notice that $630\, \mathrm{MeV} \lesssim \Sigma_{12} \lesssim 710\,\mathrm{MeV}$. 
If we assume that $400\,\mathrm{MeV} \lesssim m_{B^*_1}-m_B \lesssim 500\,\mathrm{MeV}$ in 
the pion mass range $[310\, \mathrm{MeV},\, 440\, \mathrm{MeV}]$, (as has been found in
a recent lattice study of the static light meson spectrum \cite{Michaelaa}), we
conclude that our analysis is safe from these threshold effects. Moreover
the bare couplings $g_{12}$ we obtain are similar to the quenched result of
Ref. \cite{BulavaYZ}.

\begin{figure*}[t!]
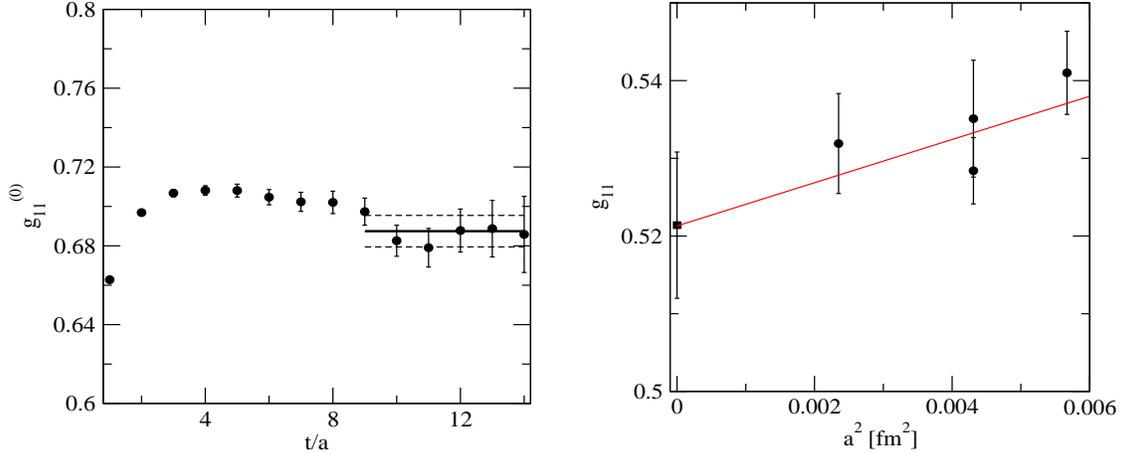

\includegraphics*[width=7cm, height=6cm]{g11sGEVP.eps} \quad\quad
\includegraphics*[width=7cm, height=6cm]{g11extrcont.eps}\\
\caption{\label{figg11} Plateau of bare $g_{11}$ for the CLS ensemble E5 (left)
and its extrapolation to the continuum and chiral limit (right).}
\end{figure*}

\begin{figure*}[t!]
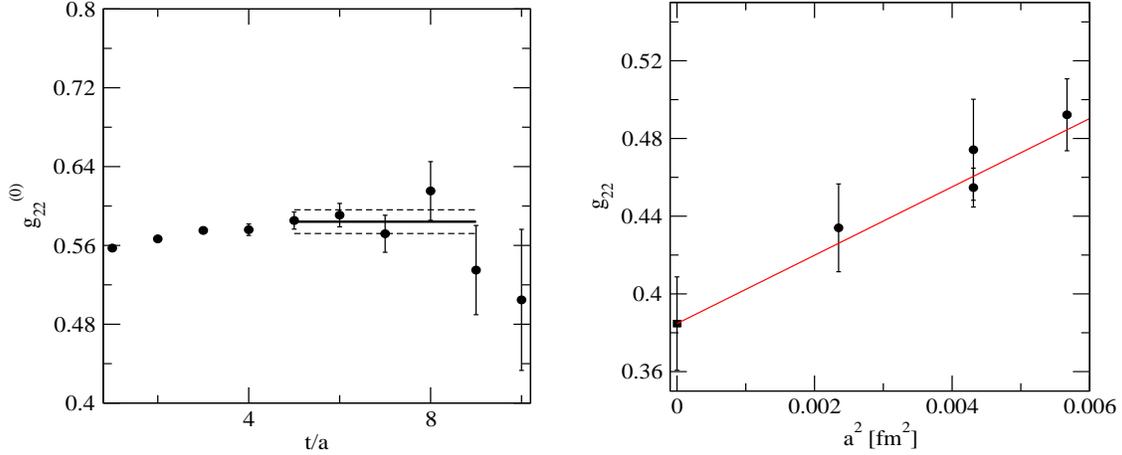

\includegraphics*[width=7cm, height=6cm]{g22sGEVP.eps} \quad\quad
\includegraphics*[width=7cm, height=6cm]{g22extrcont.eps}\\
\caption{\label{figg22} Plateau of bare $g_{22}$ for the CLS ensemble E5 (left)
and its extrapolation to the continuum and chiral limit (right).}
\end{figure*}

 We show in Figure \ref{figg11} a typical plateau of the bare coupling 
 $g_{11}$ and the extrapolation to the continuum and chiral limit. 
That extrapolation is smooth, with a negligible dependence on $m_\pi$, and we 
obtain from the fit form (\ref{fit1})
$g_{11}=0.52(2)$, in excellent agreement with a computation by the ALPHA 
Collaboration focused on 
that quantity~\cite{BulavaEJ}. 
We have added an error of 2\% due to higher excited states which is 
estimated from plateaux at early times with a range
ending at $\sim r_0$.
Following the same 
strategy, we show in Figure~\ref{figg22} 
a typical plateau of the bare coupling $g_{22}$ and the extrapolation to the 
continuum and chiral limit, once again quite smooth, with an almost absent 
dependence on the sea quark mass. 
We obtain from the fit form (\ref{fit2}) $g_{22}=0.38(4)$. Remarkably, the 
``diagonal" couplings 
$g_{11}$ and $g_{22}$ are significantly larger than the off-diagonal one $g_{12}$. 
This suggests that neglecting the contribution from $B'$ mesons to the 
three-point light-cone sum rule used to obtain $g_{B^* B\pi}$ introduces 
uncontrolled systematics. 
Note that the decay constant $f_{B^{*'}}$ itself is large compared to 
$f_B$~\cite{BurchQX,BlossierMK}.
For completeness we have collected in Table~\ref{tabg1122} the value of 
$g_{11}$ and $g_{22}$ at each 
lattice point and at the physical point and 
the fit parameters of (\ref{fit1}) and (\ref{fit2}).
\begin{table}
\begin{center}
\begin{tabular}{|c|c|c|}
\cline{2-3}
\multicolumn{1}{l|}{}&$g_{11}$&$g_{22}$\\
\hline
A5&0.541(5)&0.492(19)\\
E5&0.535(8)&0.455(10)\\
F6&0.528(4)&0.474(26)\\
N6&0.532(6)&0.434(23)\\
\hline
physical point&0.516(12)(5)(10)&0.385(24)(28)\\
\hline
\end{tabular}
\quad
\begin{tabular}{|c|c|c|c|c|}
\cline{2-5}
\multicolumn{1}{l|}{}&$g_{11}$: fit (\ref{fit1})&$g_{11}$: fit (\ref{fit2})
&$g_{22}$: fit (\ref{fit1})&$g_{22}$: fit (\ref{fit2})\\
\hline
$C_{0}$&0.515(13)&0.521(9)&0.416(27)&0.385(24)\\
$C_{1}$&0.012(9)&0.012(9)&0.074(25)&0.076(26)\\
$C_{2}$&0.0011(15)&-&-0.0033(33)&-\\
\hline
\end{tabular}
\end{center}
\caption{\label{tabg1122} Value of $g_{11}$ and $g_{22}$ at the lattice points and at the 
physical point (left) and fit parameters of eq. (\ref{fit1}) and (\ref{fit2}) (right). The third error on $g_{11}$ is
an estimate of the effects of higher excited states.}
\end{table}

\section{\label{sec4}Conclusion}

We have performed a first estimate of the axial form factor
$A_1(q^2_{\rm max})\equiv g_{12}$ parametrising at zero recoil the decay $B^{*'} \to B$ in the static 
limit of HQET from ${\rm N_f=2}$ lattice simulations. Assuming the positivity 
of decay constants $f_B$ and $f_{B^{*'}}$, we have obtained a 
\emph{negative} value for this
form factor. It is almost three times smaller than the $g_{11}$ coupling: $g_{12}=-0.17(4)$ while 
$g_{11}=0.52(2)$. Moreover we find $g_{22}=0.38(4)$, which is not strongly 
suppressed with respect to $g_{11}$. Our work is a first hint of confirmation of the statement made in 
Ref.~\cite{BecirevicVP} to explain the
small value of $g_{D^*D\pi}$ computed analytically when compared to experiment. This computation using light-cone Borel sum rules 
may have been too naive. Following Ref.~\cite{BecirevicYA}, a next step in our 
general study of excited static-light meson states would be the measurement 
of $A_1(0)$ by computing the distribution in $r$ of the axial density $f_A(r)
\equiv \elematrice{B^{*'}}{\bar{\psi}_l\gamma^i\gamma^5 \psi_l(r)}{B}$ and 
$A_1(0)=4\pi\,\int_0^\infty r^2 f_A(r) e^{i\vec{q}\cdot \vec{r}} \,dr$.

\section*{Acknowledgements}

We thank Damir Becirevic, Nicolas Garron, Alain Le Yaouanc and Rainer Sommer 
for valuable discussions and our colleagues in 
the CLS effort for the use of gauge configurations.
B. B. thanks the Galileo Galilei Institute for 
Theoretical Physics for the hospitality and the INFN for partial support 
during the completion of 
this work. Work by M. D. has been supported by the EU Contract 
No.~MRTN-CT-2006-035482, ``FLAVIAnet''. Computations of the
relevant correlation functions are made on GENCI/CINES,
under the Grants 2012-056806 and 2013-056806.

\section*{Appendix}

In this section we discuss the time dependence of $R^{\rm sGEVP}_{mn}$ (\ref{ratiosgevp}). To simplify notation, we have fixed $a$ to 1. We have followed the 
strategy of Ref.~\cite{BlossierKD} to treat in perturbation theory the full 
GEVP, with an exact computation of the $N$ lowest states:
\begin{align*}
&C^{(2)}_{ij}(t) = \langle \mathcal{O}_i(t) \mathcal{O}_j(0) \rangle = C_{ij}^{(2,0)}(t) + \epsilon C_{ij}^{(2,1)}(t) =  \sum_{n=1}^N e^{-E_nt} \psi_{ni} \psi_{nj}  +  \sum_{n=N+1}^{\infty} e^{-E_nt} \psi_{ni} \psi_{nj}\,, \\
&C^{(2)}(t)v_n(t,t_0) = \lambda_n(t,t_0) C^{(2)}(t_0) v_n(t,t_0)\,,\\
&v_n(t,t_0) = v_n^{(0)}(t,t_{0}) + \epsilon v_n^{(1)}(t,t_0)\,,  \\
&\lambda_n(t,t_0) = \lambda_n^{(0)}(t,t_{0}) + \epsilon \lambda_n^{(1)}(t,t_0)\,.
\end{align*} 
Vectors are normalised such that
\begin{align*}
& \langle v_m^{(0)},C^{(2,0)}(t_0) v_n^{(0)} \rangle= \rho_n \delta_{nm}\,,\\
& \langle v_n^{(1)},C^{(2,0)}(t_0) v_n^{(0)}\rangle = 0\,.\\
\end{align*}
where $\rho_n=e^{-E_nt_0}$. Introducing the dual vectors $u_{n}$ defined by 
$\sum_{n=1}^{N} u_{ni} \psi_{mi}  = \delta_{mn}$ $\forall n \leq N$, we note 
that
\beq\nonumber
C^{(2,0)}(t)u_n = e^{-E_n t}  \ \psi_{n}, \quad v_n^{(0)}(t,t_0) = u_n, \quad 
\lambda_n^{(0)}(t,t_0) = e^{-E_n(t-t_0)}.
\eeq
At first order in $\epsilon$, we have
\begin{align*}
 \lambda_n^{(1)} &= \rho_n^{-1} \left(v_n^{(0)}, \Delta_n v_n^{(0)} \right)\,, \\
 v_n^{(1)} &= \sum_{m \neq n} v_m^{(0)} \rho_m^{-1} \frac{ (v_m^{(0)}, \Delta_n v_n^{(0)})}{\lambda_n^{(0)} - \lambda_m^{(0)}} = \sum_{n \neq m} \alpha_{nm} \ v_m^{(0)},
\end{align*}
where $\Delta_n = C^{(2,1)}(t) - \lambda^{(0)}_n(t,t_{0}) C^{(2,1)}(t_0)$. With
$c_{n,m,l} = \langle u_n,\psi_l\rangle \langle u_m,\psi_l\rangle $ we get:
\begin{align*}
\epsilon \frac{\lambda_n^{(1)}(t,t_0)}{\lambda_n^{(0)}(t,t_0)} &= - \sum_{l>N} c_{n,n,l}  e^{-(E_l-E_n)t_0} \left[ 1 -  e^{-(E_{l}-E_n)(t-t_0)} \right]\,, \\
\epsilon \ \alpha_{nm}(t,t_0) &= - \sum_{l>N}   c_{n,m,l} \frac{1- e^{-(E_l-E_n)(t-t_0)}}{1 - e^{-(E_m-E_n)(t-t_0)}}  e^{-(E_l-E_m)t_0} \,.
\end{align*}
Finally the normalisation conditions read
\beq\nonumber
\langle v_n(t,t_0),C(t_0) v_n(t,t_0) \rangle= \rho_n + \epsilon \langle v_n^{(0)}, C^{(2,1)}(t_0) v_n^{(0)} \rangle \,.
\eeq
We are ready to develop (\ref{ratiosgevp}) to first order in $\epsilon$:
\bea\nonumber
\mathcal{M}_{mn}^{\mathrm{eff},s}& =& \partial_t \left\{  \frac{\langle v_m(t,t_0),   \left[ K(t,t_0) / \lambda_n(t,t_0) -  K(t_0,t_0) \right] v_n(t,t_0)  \rangle}{\left[  \langle v_m(t,t_0), C^{(2)}(t_0)v_m(t,t_0)  \rangle 
\langle v_n(t,t_0),C^{(2)}(t_0)v_n(t,t_0)  \rangle  \right]^{1/2} } e^{ \frac{t_0}{2} \Sigma(t_0,t_0)} \right\} \\
\nonumber
&=&\mathcal{M}_{mn}^{\mathrm{eff},s,0} + \epsilon \mathcal{M}_{mn}^{\mathrm{eff},s,1},
\eea
\begin{align*}
\mathcal{M}_{mn}^{\mathrm{eff},s,0} &= \partial_t \left\{ \frac{ \langle u_m,   \left[ K(t,t_0) / \lambda_n^{(0)}(t,t_0) -  K(t_0,t_0) \right] u_n  \rangle }{  \left[  \langle u_m, C^{(2)}(t_0)u_m \rangle 
\langle u_n,C^{(2)}(t_0)u_n \rangle  
\right]^{1/2}  }  e^{ \frac{t_0}{2} \Sigma(t_0,t_0)}  \right\}.
\end{align*}
 With
\beq\nonumber
\langle u_m,  K(t,t_0) u_n  \rangle  = \sum_{t_1} e^{-\Sigma (t-t_1)} 
\langle u_m,  C^{(3)}(t,t_1) u_n  \rangle =  h_{mn}\ t e^{-E_n t},\quad
\langle u_n,  C^{(2)}(t_0) u_n  \rangle = \rho_n = e^{-E_n t_0}
\eeq
we have at leading order
\beq\nonumber
\mathcal{M}_{mn}^{\mathrm{eff},s,0} = h_{mn}.
\eeq
The subleading order reads
\beq\nonumber
\epsilon \mathcal{M}_{mn}^{\mathrm{eff},s,1}=\epsilon \partial_{t} \sum_{a=1}^{5} T_{a}.
\eeq
\beq\nonumber
T_{1}=-\frac{\lambda_n^{(1)}(t,t_0)}{( \lambda_n^{(0)}(t,t_0) )^2} \frac{\langle v_m^{(0)}(t,t_0),   K(t,t_0)  v_n^{(0)}(t,t_0)  \rangle}{(\rho_n \rho_m)^{1/2} }  e^{ \frac{t_0}{2} \Sigma(t_0,t_0)} = -\frac{\lambda_n^{(1)}(t,t_0)}{\lambda_n^{(0)}(t,t_0)}\langle v_m^{(0)}(t,t_0),   K(t,t_0)  v_n^{(0)}(t,t_0)  \rangle e^{E_nt}\,.
\eeq
The first subleading contribution is given by
\begin{align*}
T_1 =  - h_{mn}\ t \times \frac{\lambda_n^{(1)}(t,t_0)}{\lambda_n^{(0)}(t,t_0)}  \sim c_{n,n,N+1} \ h_{mn}\times t e^{-\Delta_{N+1,n}t_0} \left[ 1 -  e^{-\Delta_{N+1,n}(t-t_0)} \right]\,. 
\end{align*}
Defining the discrete derivative $\partial_t A = A(t+1)-A(t)$, and taking at the end of the computation
$t_{0}=t-1$, we get
\begin{align*}
\partial_t T_1 \sim c_{n,n,N+1} \ h_{mn} \left( 1 - e^{-\Delta_{N+1,n}} \right) \times \left(t+ 1 + e^{\Delta_{N+1,n}} \right)  e^{-\Delta_{N+1,n} t}.
\end{align*}
The second subleading contribution reads
\begin{multline*}
T_{2}=\frac{\left( v_m^{(1)}(t,t_0),   \left[ K(t,t_0) / \lambda_n^{(0)}(t,t_0) - K(t_0,t_0) \right]  v_n^{(0)}(t,t_0)  \right)}{(\rho_n \rho_m)^{1/2}}  e^{ \frac{t_0}{2} \Sigma(t_0,t_0)} = \\
\sum_{p \neq m} \alpha_{mp}\ \langle v_p^{(0)}(t,t_0), \left[  K(t,t_0)e^{E_nt} - K(t_0,t_0)e^{E_nt_0} \right]  v_n^{(0)}(t,t_0)  \rangle\,. 
 \end{multline*}
With some algebra, we deduce
\begin{flalign*}
\langle v_p^{(0)}(t,t_0), K(t,t_0) v_n^{(0)}(t,t_0) \rangle e^{E_n t}  &= \sum_{t_1} e^{-(t-t_1)(E_n-E_m)} \sum_{rs} \langle u_p,\psi_{r}\rangle \langle \psi_{s},u_n\rangle h_{rs} e^{-E_r(t-t_1)} e^{-E_s t_1} e^{E_n t}&\\
&= \sum_{t_1} e^{-(t-t_1)(E_n-E_m)} h_{pn} e^{-E_p(t-t_1)} e^{-E_n t_1} e^{E_n t}&\\
&= \sum_{t_1}  h_{pn} e^{-(E_p-E_m)t_1},&
\end{flalign*}
and
\begin{flalign*}
\langle v_p^{(0)}(t,t_0),\left[K(t,t_0) e^{E_n t} - K(t_0,t_0) e^{E_n t_0} \right]  v_n^{(0)}(t,t_0) \rangle &= \sum_{t_1=t_0+1}^t  h_{pn} e^{-(E_p-E_m)t_1}.&
\end{flalign*}
Finally,
\begin{align*}
T_2 &= \sum_{p \neq m} \left[ \alpha_{mp}(t,t_0)  \sum_{t_1=t_0+1}^t h_{pn} e^{-(E_p-E_m)t_1} \right] , &
\end{align*}
\begin{align*}
\partial_t T_2 = \sum_{p \neq m}  \left[    \left( \alpha_{mp}(t+1,t_0)-\alpha_{mp}(t,t_0) 
\right) \sum_{t_1=t_0+1}^t h_{pn} e^{-(E_p-E_m)t_1} +     \alpha_{mp}(t+1,t_0) \ h_{pn} e^{-(E_p-E_m)(t+1)}   
\right]\,.
\end{align*}
Setting $t_{0}=t-1$, the first term reads
\begin{multline*}
\sum_{p \neq m} \left( \alpha_{mp}(t+1,t_0)-\alpha_{mp}(t,t_0) \right) \times e^{-(E_p-E_m)t} \\
 \sim - \sum_{p \neq m} \left[ c_{m,p,N+1} e^{-(E_{N+1}-E_p)(t-1)} \times \left( \frac{1-e^{-2(E_{N+1}-E_m)}}{1-e^{-2(E_p-E_m)}} - \frac{1-e^{-(E_{N+1}-E_m)}}{1-e^{-(E_p-E_m)}} \right)\right] \times h_{pn} e^{-(E_p-E_m)t} \\
\sim - e^{-(E_{N+1}-E_m)t} \sum_{p \neq m} \left[ c_{m,p,N+1} h_{pn} e^{(E_{N+1}-E_p)} \times \left( \frac{1-e^{-2(E_{N+1}-E_m)}}{1-e^{-2(E_p-E_m)}} - \frac{1-e^{-(E_{N+1}-E_m)}}{1-e^{-(E_p-E_m)}} \right)\right]\,,  
\end{multline*}
and the second term reads
\begin{multline*}
\sum_{p \neq m}  \alpha_{mp}(t+1,t_0) \ h_{pn} e^{-(E_p-E_m)(t+1)} \\
\sim - \sum_{p \neq m} e^{-(E_{N+1}-E_p)(t-1)}  \frac{1-e^{-2(E_{N+1}-E_m)}}{1-e^{-2(E_p-E_m)}} c_{m,p,N+1} \ h_{pn} \times e^{-(E_p-E_m)(t+1)} \\
\sim - e^{-(E_{N+1}-E_m)t} \sum_{p \neq m} e^{(E_{N+1}+E_m-2E_p)} \frac{1-e^{-2(E_{N+1}-E_m)}}{1-e^{-2(E_p-E_m)}} c_{m,p,N+1} \ h_{pn}.
\end{multline*}
We find
\begin{equation*}
\partial_t T_2 \sim\  e^{-(E_{N+1}-E_m)t} \sum_{p \neq m} c_{m,p,N+1} h_{pn}  \frac{ 1 - e^{-(E_{N+1}-E_m)}}{1-e^{-(E_m-E_p)}} \,.
\end{equation*}
The third contribution
\begin{align*}
T_{3}=\frac{\left( v_m^{(0)}(t,t_0),   \left[ K(t,t_0) / \lambda_n^{(0)}(t,t_0) - K(t_0,t_0) \right]  v_n^{(1)}(t,t_0)  \right)}{(\rho_n \rho_m)^{1/2}}  e^{ \frac{t_0}{2} \Sigma(t_0,t_0)}
 \end{align*}
is obtained similarly to $\partial_{t}T_{2}$, permuting $m$ and $n$.\\
The fourth subleading contribution reads
\bea
\nonumber T_{4}&=& \frac{1}{\lambda_n^{(0)}(t,t_0)} \frac{\langle v_m^{(0)}(t,t_0),   K^{(1)}(t,t_0)  v_n^{(0)}(t,t_0)  \rangle}{(\rho_n \rho_m)^{1/2} }  e^{ \frac{t_0}{2} \Sigma(t_0,t_0)} = \langle v_m^{(0)}(t,t_0),   K^{(1)}(t,t_0)  v_n^{(0)}(t,t_0)  \rangle e^{E_nt}\,.
\eea
With some algebra we deduce 
\begin{align*}
\langle v_m^{(0)}(t,t_0),K^{(1)}(t,t_0) v_n^{(0)}(t,t_0) \rangle &= \sum_{t_1} e^{-(E_n-E_m)(t-t_1)} 
\sum_{(r\, {\rm or}\, s) > N} \langle u_m,\psi_{r}\rangle \langle \psi_{s},u_n\rangle h_{rs} e^{-E_r(t-t_1)} e^{-E_s t_1}\\
&\hspace{1cm} =  \sum_{t_1} e^{-(E_n-E_m)(t-t_1)} \langle u_m,\psi_{N+1}\rangle h_{N+1,n} e^{-E_{N+1}(t-t_1)} e^{-E_n t_1} \\
&\hspace{1cm} + \sum_{t_1} e^{-(E_n-E_m)(t-t_1)} \langle u_n,\psi_{N+1}\rangle  h_{N+1,m} e^{-E_{m}(t-t_1)} e^{-E_{N+1} t_1} \\
&\hspace{1cm} + \sum_{t_1} e^{-(E_n-E_m)(t-t_1)} \sum_{(r, s) > N}  \langle u_n,\psi_{r}\rangle \langle u_m,\psi_{s}\rangle h_{r,s} e^{-E_r(t-t_1)} e^{-E_s t_1} \\
& \hspace{2cm} \sim \sum_{t_1} e^{-(E_n-E_m)t_1} \langle u_m,\psi_{N+1}\rangle h_{N+1,n} e^{-E_{N+1}t_1} e^{-E_n (t-t_1)} \\ 
& \hspace{2cm} + \sum_{t_1} e^{-E_n(t-t_1)} \langle u_n,\psi_{N+1}\rangle h_{N+1,m}  e^{-E_{N+1} t_1} \\
& \hspace{2cm} + \sum_{t_1} e^{-(E_n-E_m)(t-t_1)} \langle u_n,\psi_{N+1}\rangle \langle u_m,\psi_{N+1}\rangle h_{N+1,N+1} e^{-E_{N+1}t}\\
& \hspace{3cm} \sim  e^{-E_n t}  \langle u_m,\psi_{N+1}\rangle h_{N+1,n} \sum_{t_1}  e^{-(E_{N+1}-E_m)t_1}  \\
& \hspace{3cm} + e^{-E_n t} \langle u_n,\psi_{N+1}\rangle h_{N+1,m}  \sum_{t_1} e^{-(E_{N+1} -E_n) t_1} \\
& \hspace{3cm} +c_{n,m,N+1} \ h_{N+1,N+1} \ e^{-E_{N+1}t} \sum_{t_1} e^{-(E_n-E_m)t_1}\,,  
\end{align*}
and we obtain
\begin{align*}
\partial_t T_4 \sim &+\langle u_m,\psi_{N+1}\rangle  h_{N+1,n}  e^{-(E_{N+1}-E_m)(t+1)} \\
& + \langle u_n,\psi_{N+1}\rangle  h_{N+1,m} e^{-(E_{N+1} -E_n) (t+1)}\\
& - c_{n,m,N+1} h_{N+1,N+1}  \frac{  e^{-(E_{N+1}-E_n)} - 1 }{e^{-(E_n-E_m)}-1}  e^{-(E_{N+1}-E_n)t}  \\
& - c_{n,m,N+1} h_{N+1,N+1}  \frac{  e^{-(E_{N+1}-E_m)} - 1 }{e^{-(E_m-E_n)}-1}  e^{-(E_{N+1}-E_m)t}\,. 
\end{align*}
The last subleading contribution reads
\bea\nonumber
T_{5}&=& - t \ h_{mn}  \times \left(  \frac{\langle v_m^{(0)},C^{(2,1)}(t_0)v_m^{(0)}\rangle}{2\rho_m} +  \frac{\langle v_n^{(0)},C^{(2,1)}(t_0)v_n^{(0)}\rangle} {2\rho_n}\right) \\\nonumber
&\sim&- t \ h_{mn} \times \left( \frac{1}{2} c_{m,m,N+1} e^{-(E_{N+1}-E_m)t_0} + \frac{1}{2} c_{n,n,N+1} e^{-(E_{N+1}-E_n)t_0}  \right).
\eea
With $t_{0}=t-1$, we get
\beq\nonumber
\partial_{t} T_5 \sim - \frac{h_{mn}}{2} \times \left( c_{m,m,N+1} e^{-(E_{N+1}-E_m)(t-1)} +  c_{n,n,N+1} e^{-(E_{N+1}-E_n)(t-1)}  \right)
\eeq
We see that for $n > m$ the dominating contribution $T_{1}$ to $\epsilon 
M^{{\rm eff}, s, 1}_{mn}$ is in $t e^{-\Delta_{N+1,n} t}$ with subleading terms 
$T_{2} - T_{5}$ while for $n < m$ the leading contribution is in 
$e^{-(E_{N+1}-E_m)t}$.

\begin{figure}[h!]
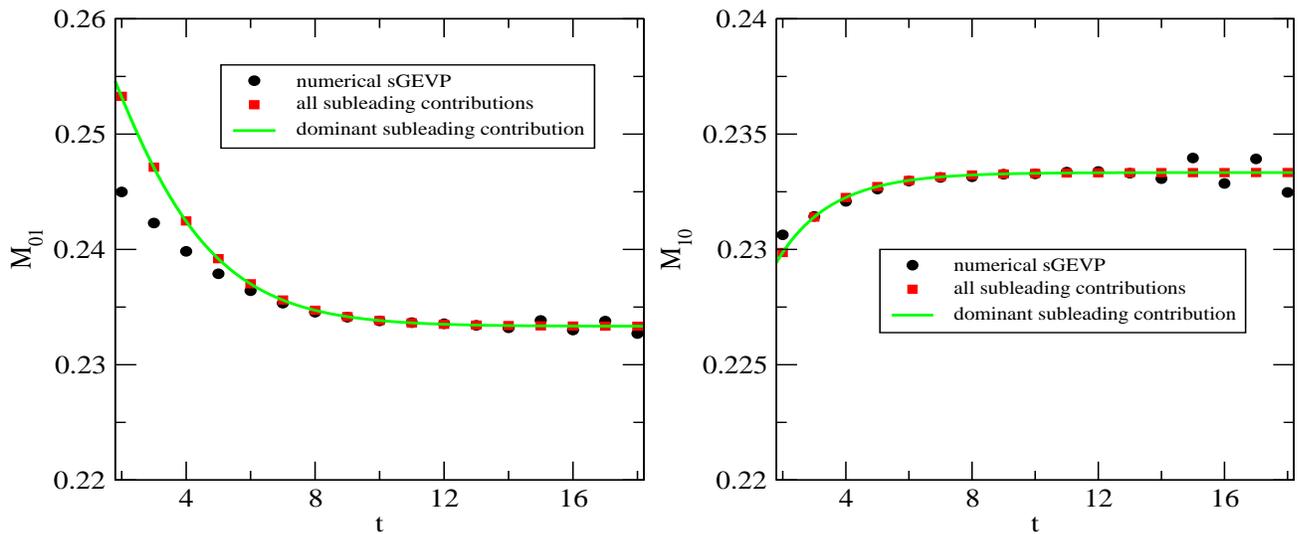

\begin{center}
\begin{tabular}{cc}
\includegraphics*[width=8.5cm,height=7cm]{sGEVPtoymodel01.eps}
&
\includegraphics*[width=8.5cm,height=7cm]{sGEVPtoymodel10.eps}
\end{tabular}
\end{center}
\caption{\label{figcompsgevp} Analytical formulae for $R^{\rm sGEVP}_{mn}$ compared to the numerical
solution of our toy model.}
\end{figure}

\vspace{0.4cm}
We have tested numerically our finding in the toy model of 
Ref.~\cite{BulavaYZ}, with $r_{0} E_{n}=n$, 
$r_{0}=0.3$, the $3 \times 5$ matrix of couplings
\beq\nonumber
\psi = \langle 0 | \mathcal{O}_i | n\rangle = \begin{pmatrix}
0.92 & 0.03 & -0.10 & -0.01 & -0.02\\
0.84 & 0.40 & 0.03 & -0.06 & 0.00 \\
0.56 & 0.56 & 0.47 & 0.26 & 0.04
\end{pmatrix},
\eeq
and the hadronic matrix elements 
$M_{nn}=0.7 \frac{6}{n+5}$, $M_{n,n+m}=\frac{M_{nn}}{3m}$.
The comparison between the analytical formulae and the numerical solution is 
plotted in Figure \ref{figcompsgevp}. It is encouraging to obtain such good 
agreement after $t=8$.

\bibliographystyle{style.bst}
\bibliography{cited_refs.bib}

\begin{thebibliography}{10}

\bibitem{GambinoRD}
P.~Gambino, T.~Mannel, and N.~Uraltsev,
\newblock JHEP {\bf 1210}, 169 (2012), 1206.2296.

\bibitem{BecirevicVP}
D.~Becirevic {\em et~al.},
\newblock JHEP {\bf 0301}, 009 (2003), hep-ph/0212177.

\bibitem{OhkiPY}
H.~Ohki, H.~Matsufuru, and T.~Onogi,
\newblock Phys.Rev. {\bf D77}, 094509 (2008), 0802.1563.

\bibitem{BecirevicYB}
D.~Becirevic, B.~Blossier, E.~Chang, and B.~Haas,
\newblock Phys.Lett. {\bf B679}, 231 (2009), 0905.3355.

\bibitem{BulavaEJ}
ALPHA Collaboration, J.~Bulava, M.~Donnellan, and R.~Sommer,
\newblock PoS {\bf LATTICE2010}, 303 (2010), 1011.4393.

\bibitem{DetmoldGE}
W.~Detmold, C.~D. Lin, and S.~Meinel,
\newblock Phys.Rev. {\bf D85}, 114508 (2012), 1203.3378.

\bibitem{BecirevicPF}
D.~Becirevic and F.~Sanfilippo,
\newblock Phys.Lett. {\bf B721}, 94 (2013), 1210.5410.

\bibitem{BernardoniIP}
ALPHA Collaboration, F.~Bernardoni, J.~Bulava, M.~Donnellan, and R.~Sommer,
\newblock (In preparation).

\bibitem{Godang:2013im}
R.~Godang,
\newblock (2013), 1301.0141.

\bibitem{KhodjamirianHB}
A.~Khodjamirian, R.~Ruckl, S.~Weinzierl, and O.~I. Yakovlev,
\newblock Phys.Lett. {\bf B457}, 245 (1999), hep-ph/9903421.

\bibitem{BulavaNP}
J.~Bulava,
\newblock PoS {\bf LATTICE2011}, 021 (2011), 1112.0212.

\bibitem{BurchCC}
T.~Burch {\em et~al.},
\newblock Phys.Rev. {\bf D74}, 014504 (2006), hep-lat/0604019.

\bibitem{BlossierVZ}
ALPHA Collaboration, B.~Blossier {\em et~al.},
\newblock JHEP {\bf 1005}, 074 (2010), 1004.2661.

\bibitem{MohlerKE}
D.~Mohler and R.~Woloshyn,
\newblock Phys.Rev. {\bf D84}, 054505 (2011), 1103.5506.

\bibitem{MahbubRM}
CSSM Lattice collaboration, M.~S. Mahbub, W.~Kamleh, D.~B. Leinweber, P.~J.
  Moran, and A.~G. Williams,
\newblock Phys.Lett. {\bf B707}, 389 (2012), 1011.5724.

\bibitem{MichaelNE}
C.~Michael,
\newblock Nucl.Phys. {\bf B259}, 58 (1985).

\bibitem{LuscherCK}
M.~Luscher and U.~Wolff,
\newblock Nucl.Phys. {\bf B339}, 222 (1990).

\bibitem{BlossierKD}
B.~Blossier, M.~Della~Morte, G.~von Hippel, T.~Mendes, and R.~Sommer,
\newblock JHEP {\bf 0904}, 094 (2009), 0902.1265.

\bibitem{BulavaYZ}
J.~Bulava, M.~Donnellan, and R.~Sommer,
\newblock JHEP {\bf 1201}, 140 (2012), 1108.3774.

\bibitem{HasenfratzHP}
A.~Hasenfratz and F.~Knechtli,
\newblock Phys.Rev. {\bf D64}, 034504 (2001), hep-lat/0103029.

\bibitem{DellaMorteYC}
M.~Della~Morte, A.~Shindler, and R.~Sommer,
\newblock JHEP {\bf 0508}, 051 (2005), hep-lat/0506008.

\bibitem{FoleyAC}
J.~Foley {\em et~al.},
\newblock Comput.Phys.Commun. {\bf 172}, 145 (2005), hep-lat/0505023.

\bibitem{GuskenAD}
S.~Gusken {\em et~al.},
\newblock Phys.Lett. {\bf B227}, 266 (1989).

\bibitem{AlbaneseDS}
APE Collaboration, M.~Albanese {\em et~al.},
\newblock Phys.Lett. {\bf B192}, 163 (1987).

\bibitem{DellaMorteXB}
M.~Della~Morte, R.~Sommer, and S.~Takeda,
\newblock Phys.Lett. {\bf B672}, 407 (2009), 0807.1120.

\bibitem{FritzschWQ}
P.~Fritzsch {\em et~al.},
\newblock Nucl.Phys. {\bf B865}, 397 (2012), 1205.5380.

\bibitem{CasalbuoniPG}
R.~Casalbuoni {\em et~al.},
\newblock Phys.Rept. {\bf 281}, 145 (1997), hep-ph/9605342.

\bibitem{BurdmanGH}
G.~Burdman and J.~F. Donoghue,
\newblock Phys.Lett. {\bf B280}, 287 (1992).

\bibitem{SommerCE}
R.~Sommer,
\newblock Nucl.Phys. {\bf B411}, 839 (1994), hep-lat/9310022.

\bibitem{Michaelaa}
ETM Collaboration, C.~Michael, A.~Shindler, and M.~Wagner,
\newblock JHEP {\bf 1008}, 009 (2010), 1004.4235.

\bibitem{BurchQX}
T.~Burch, C.~Hagen, C.~B. Lang, M.~Limmer, and A.~Schafer,
\newblock Phys.Rev. {\bf D79}, 014504 (2009), 0809.1103.

\bibitem{BlossierMK}
ALPHA Collaboration, B.~Blossier {\em et~al.},
\newblock JHEP {\bf 1012}, 039 (2010), 1006.5816.

\bibitem{BecirevicYA}
D.~Becirevic, E.~Chang, and A.~L. Yaouanc,
\newblock Phys.Rev. {\bf D80}, 034504 (2009), 0905.3352.

\end{thebibliography}

\end{document}